\documentclass[secnumarabic,amsmath,amssymb,showpacs,nobibnotes,aps, prl,12pt]{revtex4-1}

\usepackage{amssymb}

\usepackage{graphicx}
\usepackage{dcolumn}
\usepackage{bm}
\usepackage{color}

\begin{document}
\title{Parallel acceleration due to the radial electric field in a magnetized plasma with low-frequency turbulence}

\author{Shaojie Wang}
\email{wangsj@ustc.edu.cn}
\affiliation{Department of Modern Physics, University of Science and Technology of China, Hefei, 230026, China}
\date{\today}

\begin{abstract}
A new physical mechanism of the parallel acceleration of a turbulent magnetized plasma is discovered by using a Fokker-Planck phase space stochastic transport equation. It is found that the random walk of a charged particle is correlated with the random change of the parallel velocity due to the radial electric field and the magnetic moment conservation. This correlation leads to a parallel acceleration of the plasma with a finite parallel fluid flow.
\end{abstract}

\pacs{52.25.Dg, 52.25.Fi, 52.20.Dq, 52.65.-y}

\maketitle
(I)INTRODUCTION

In a tokamak fusion experiment, the toroidal rotation is important, since it is related to the $\bm E \times \bm B$ shearing that can stabilize turbulence and thereby improve the confinement \cite{BiglaryPF90}. Intrinsic rotation without external momentum injection has been widely observed in tokamak experiments \cite{RiceNF98,RiceNF04,AngioniNF12}. To understand the intrinsic rotation, which is especially important for a tokamak fusion reactor where an external momentum injection is difficult, various theoretical models have been proposed. Momentum pinch, which was first pointed out in Ref. \onlinecite{LeePRL03}, can be generated by the Coriolis force \cite{PeetersPRL07}. Another candidate explanation of the momentum pinch is the residual parallel Reynolds stress based on the quasilinear theory \cite{DiamondPoP08}. However, it may be not sufficient to interpret the intrinsic rotation by a momentum pinch without a momentum source at the edged of plasma. The intrinsic rotation has become currently a topic of intensive researches \cite{AngioniNF12,PeetersPRL07,DiamondPoP08,GaoPRL13,WangPRL13}. The ponderomotive force due the the radio-frequency waves \cite{GaoPRL13} and the parallel acceleration due to the correlation between the turbulent fluctuations of density and ion temperature \cite{WangPRL13} have been proposed to explain the parallel acceleration.
In addition to the momentum pinch, particle pinch has also been observed in tokamak fusion experiments for many years \cite{WarePRL70, AngioniNF12}. Anomalous trapped electron pinch in a low-frequency electrostatic turbulence has been identified \cite{IsichenkoPRL95,BakerPoP98} by invoking the turbulence equipartition model, which can be understood by considering the correlation between the radial position change and the kinetic energy change induced by the $\nabla B$ force.

In this paper, we propose a new theoretical model to interpret the turbulence parallel acceleration by using a Fokker-Planck phase space stochastic transport equation, which is similar to the Fokker-Planck equation used by Baker-Rosenbluth in discussing teh anomalous particle pinch \cite{BakerPoP98}. It is found that due to the radial electric field (REF), the random change of the kinetic energy of a charged particle in a turbulent magnetized plasma is correlated with the radial random-walk; due to the magnetic moment conservation, the random change of the kinetic energy leads to a random change of the parallel velocity; therefore, the radial random walk is correlated with the random change of the parallel velocity. This correlation reveals a new physical mechanism of parallel acceleration of the turbulent magnetized plasma.

(II)TRANSPORT EQUATION 

Consider the radial random walk due to the nonlinear stochastic scattering by the low-frequency turbulence. Here, low-frequency means that $\left(\omega_{wave},1/\tau_c\right)\ll\Omega$, with $\omega_{wave}$ the characteristic frequency of the wave, $\tau_c$ the correlation time of the turbulence, and $\Omega$ the gyro-frequency of the charged particle.
Let $f_s \left(\bm z, t\right)$ denotes the ensemble averaged particle distribution in the phase space $z^i=\left(\bm x, v_{\parallel},K,\xi \right)$, with $\bm x$ the particle position, $v_{\parallel}$ the velocity component parallel to the magnetic field, $K$ the kinetic energy of the particle, $\xi$ the gyro-angle. The subscript $s=i,e$ denotes the particle species (ion, electron).
When considering the nonlinear stochastic turbulence scattering, the phase space transport equation is given by a Fokker-Planck equation \cite{WangPoP12, WangPRE13, WangPoP13},
\begin{equation}
  \partial_t  f_s =
  \frac{1}{\mathcal J}\partial_i\left\{\mathcal J\left[\frac{1}{2\tau} \left\langle G^iG^j \right\rangle _ \mathcal E \partial_j \right] f_s  \right \},\label{eq:FP}
\end{equation}
which describes the phase-space transport. $\mathcal J$ is the Jacobian of the phase-space. $\langle \rangle_{\mathcal E}$ denotes the ensemble average. Here and throughout this paper, summation over repeated numerical index is assumed.
Note that
\begin{equation}
-G^i=\Delta z^i,
\end{equation}
denotes the departure of the particle from the unperturbed orbit within a time interval $\tau$ which is longer than $\tau_c$ \cite{WangPoP12, WangPRE13, WangPoP13}, and in a tokamak the unperturbed motion does not contribute to transport by itself. Clearly, one can write
\begin{equation}
\frac{1}{2\tau} \left\langle G^iG^j \right\rangle _ \mathcal E=\frac{\left\langle \Delta z^i \Delta z^j \right\rangle _ \mathcal E}{2\Delta t}\equiv d^{ij}, \label{eq:phase-D}
\end{equation}
with $\Delta t >\tau_c$.
In a steady-state nonlinear stochastic turbulence, one has
\begin{equation}
\left\langle\Delta z^i\right\rangle_{\mathcal E}=0.\label{eq:randomr}
\end{equation}

Note that a similar Fokker-Planck equation including the correlation between the changes of particle position and kinetic energy was used by Baker-Rosenbluth in discussing the anomalous trapped particle pinch due to the inhomogeneity of the equilibrium magnetic field \cite{BakerPoP98}, and has been widely used in quasilinear transport theory \cite{ChenJGR99,KominisPRL10}.

In the following, we shall consider a slab geometry for mathematical simplicity, with $x$ the direction of inhomogeneity (the radial direction in a tokamak).
To proceed, we assume that the ensemble-averaged distribution is a local shift Maxwellian distribution $f_s\left( x,v_{\parallel},K\right)$,
\begin{equation}
 f_s=\frac{n_s}{\left(2\pi T_s/m_s\right)^{\frac{3}{2}}}exp\left(-\frac{K+\frac{1}{2}m_s U_s^2-m_sv_{\parallel}U_s}{T_s}\right),\label{eq:dist}
\end{equation}
with the averaged particle density $n_s\left( x \right)$, the temperature $T_s\left( x \right)$, the parallel fluid velocity $U_s\left(x\right)$. The plasma pressure is $p_s\left( x \right)=n_s T_s$. $m_s$ is the particle mass. $K=\frac{1}{2}m_s v_{\parallel}^2+\mu B$ is the kinetic energy of the particle, with $\mu$ the magnetic moment.

Using Eq. (\ref{eq:FP}) and Eq. (\ref{eq:phase-D}), one finds the local production rate of the entropy density,
\begin{equation}
\sigma_s= n_s \left\langle \partial_i \ln f_s d^{ij}\partial_j \ln f_s \right\rangle, \label{eq:entropy}
\end{equation}
with $\left\langle A \right\rangle \equiv \int d^3\bm v A f_s/n_s $.

Define a set of thermodynamic forces
\begin{subequations}
\begin{eqnarray}
A^s_{1}&\equiv&-\partial_x \ln p_s,\\
A^s_{2}&\equiv&-\partial_x \ln T_s,\\
A^s_{3}&\equiv&-\partial_x \ln U_s,\\
A^s_{4}&\equiv&-\partial_K \ln f_s=\frac{1}{T_s},\\
A^s_{5}&\equiv&-\partial_{v_{\parallel}} \ln f_s=-\frac{m_s U_s}{T_s},
\end{eqnarray}
\end{subequations}
where $A^s_1$, $A^s_2$, $A^s_3$ are the usual thermodynamic forces; $A^s_4$ and $A^s_5$ are two new thermodynamic forces, which represent the inhomogeneity in energy and parallel velocity.
Define the non-dimensional functions,
\begin{subequations}
\begin{eqnarray}
h^s_1&\equiv&1,\\
h^s_{2}&\equiv&\frac {K+ \frac{1}{2}m_s U_s^2-m_s v_{\parallel}U_s}{T_s}-\frac{5}{2},\\
h^s_{3}&\equiv&\frac {m_s \left(v_{\parallel}-U_s\right)U_s}{T_s}.
\end{eqnarray}
\end{subequations}

One can define the phase-space transport matrix $l^s_{ij}$, which is symmetric. Here and in the following $i,j=1,2,...5$ is understood.
\begin{equation}
l^s_{\alpha \beta}=d^{xx}h^s_{\alpha}h^s_{\beta},
\end{equation}
where and in the following $\alpha,\beta =1,2,3$ is understood.
\begin{subequations}
\begin{eqnarray}
l^s_{\alpha 4}&=&d^{xK}h^s_{\alpha},\\
l^s_{\alpha 5}&=&d^{xv_{\parallel}}h^s_{\alpha},\\
l^s_{44}&=&d^{KK},\\
l^s_{45}&=&d^{Kv_{\parallel}},\\
l^s_{55}&=&d^{v_{\parallel}v_{\parallel}}.
\end{eqnarray}
\end{subequations}

 Using the above definitions, following Eq. (\ref{eq:entropy}), one can write the production rate of the entropy density as a quadratic form
\begin{equation}
\sigma_s=A^s_{i}L^s_{ij}A^s_j,
\end{equation}
and the canonical conjugate fluxes are given by,
\begin{equation}
J^s_{i}=L^s_{ij}A^s_j,\label{eq:flux}
\end{equation}
with the macroscopic transport matrix
\begin{equation}
L^s_{ij}=n_s \langle l^s_{ij}\rangle. \label{eq:tmat}
\end{equation}

Note that the transport matrix satisfies the Onsager symmetry relation \cite{OnsagerPR31a,OnsagerPR31b}. To make use the Onsager symmetry relation, one needs the transport equations.

Integrating Eq. (\ref{eq:FP}) over the velocity space, one finds the particle transport equation
\begin{equation}
\partial_t  n_s+\partial_x \Gamma^x_s=0,\label{eq:part-t}
\end{equation}
with the radial particle flux $\Gamma_s^x$ given by
\begin{equation}
\Gamma_s^x=J^s_{1}.
\end{equation}

Taking the $\left(K+\frac{1}{2}m_s U_s^2-m_sv_{\parallel}U_s \right)$ moment of Eq. (\ref{eq:FP}), one finds the energy transport equation
\begin{equation}
\partial_t \left( \frac{3}{2}p_s\right)+\partial_x \left(q_s^x+\frac{5}{2}\Gamma_s^x T_s\right)=Q_s,\label{eq:ener-t}
\end{equation}
with $q_s^x$, the radial heat flux given by
\begin{equation}
\frac{1}{T_s}q_s^x=J^s_{2},
\end{equation}
and $Q_s$, the turbulence heating rate given by
\begin{equation}
\frac{1}{T_s}Q_s=J^s_{3}A^s_{3}+J^s_{4}A^s_{4}+J^s_{5}A^s_{5}. \label{eq:t-h}
\end{equation}

Taking the $m_s v_{\parallel}$ moment of Eq. (\ref{eq:FP}), one finds the parallel momentum transport equation
\begin{equation}
\partial_t \left( n_s m_s U_s\right)+\partial_x \left(\Pi_{\parallel s}^x+ \Gamma_s^x m_s U_s\right)=F_s,\label{eq:momentum-t}
\end{equation}
with $\Pi_{\parallel s}^x$, the radial component of the parallel viscosity given by
\begin{equation}
\frac{U_s}{T_s}\Pi_{\parallel s}^x=J^s_{3},
\end{equation}
and $F_s$, the turbulence parallel acceleration rate given by
\begin{equation}
\frac{1}{m_s}F_s=J^s_{5}. \label{eq:t-a}
\end{equation}

Note that the usual particle diffusivity, thermal conductivity, and  parallel viscosity coefficient, whose standard units are $m^2/s$, are given by $D_s=L^s_{11}/n_s$, $\chi_s=L^s_{22}/n_s$, and $\nu_s=L^s_{33}T_s/n_s m_s U_s^2$, respectively.

(III)PARALLEL ACCELERATION DUE TO THE RADIAL ELECTRIC FIELD
 
Since generally $\Gamma_i^x\neq\Gamma_e^x$ in a turbulent magnetized plasma, one needs to introduce the ambipolar REF, $E_x=-\partial_x \Phi\left(x\right)$.
Note that neoclassical collisional transport is auto-ambipolar \cite{RutherfordPF70}; a steady-state REF does not generate a radial electrical current in the framework of neoclassical transport theory. However, this auto-ambipolarity may be broken in a turbulent plasma. It has been pointed out that usually the diffusive ion flux is much larger than the electron flux, therefore, to maintain the ambipolarity (zero radial electrical current) when considering the anomalous trapped electron pinch \cite{BakerPoP98}, an ambipolar REF should be built.
Usually, $\Delta K$ and $\Delta v_{\parallel}$ do not correlate with $\Delta x$. However, the introduction of an external force, such as the ambipolar REF, makes them correlated with each other; this will be shown in the following.

A step of random walk $\Delta x$ inevitably changes the kinetic energy of the particle with charge $e_s$,
\begin{equation}
\Delta K=e_s E_x \Delta x. \label{eq:dk}
\end{equation}

Using the fact that the magnetic moment is conserved in a magnetized plasma, one finds
\begin{equation}
\Delta K=m_s v_{\parallel} \Delta v_{\parallel},\label{eq:dkdvpar}
\end{equation}
where we have used $K=\frac{1}{2}m_s v_{\parallel}^2+\mu B$, and $\partial_x B$ is ignored for simplicity. If one considers the $\partial_x B$ effect in the toroidal geometry, one finds the anomalous trapped particle pinch \cite{IsichenkoPRL95,BakerPoP98}, which shall be briefly discussed in the following. Note that here we have also assumed $\Delta v_{\parallel}\ll v_{\parallel}$.
Using Eq. (\ref{eq:dkdvpar}), one finds the change of the parallel velocity,
\begin{equation}
\Delta v_{\parallel}=e_s E_x \Delta x \frac{1}{m_s v_{\parallel}}. \label{eq:dvpar}
\end{equation}
Note that the apparent $1/v_{\parallel}$ singularity here and in the following is due to the assumption $\Delta v_{\parallel}\ll v_{\parallel}$ that we have used; this singularity can be removed by modifying Eq. (\ref{eq:dvpar}) to properly consider the small $v_{\parallel}$ limit.

With Eq. (\ref{eq:dk}) and Eq. (\ref{eq:dvpar}), one finds that $\Delta x$, $\Delta K$, and $\Delta v_{\parallel}$ are correlated with each other, therefore,
\begin{subequations}
\begin{eqnarray}
d^{xK}&=&e_s E_x d^{xx},\\
d^{KK}&=&\left(e_s E_x\right)^2 d^{xx},\\
d^{xv_{\parallel}}&=&e_s E_x d^{xx} \frac{1}{m_s v_{\parallel}},\\
d^{v_{\parallel}v_{\parallel}}&=&\left(e_s E_x\right)^2 d^{xx} \left(\frac{1}{m_s v_{\parallel}}\right)^2.
\end{eqnarray}
\end{subequations}

The effects of the ambipolar REF on the microscopic transport matrix are given by
\begin{subequations}
\begin{eqnarray}
l^s_{\alpha 4}&=&e_s E_x l^s_{11}h^s_{\alpha},\\
l^s_{\alpha 5}&=&e_s E_x l^s_{11}h^s_{\alpha}\frac{1}{m_s v_{\parallel}},\\
l^s_{44}&=&\left(e_s E_x\right)^2 l^s_{11},\\
l^s_{45}&=&\left(e_s E_x\right)^2 l^s_{11}\frac{1}{m_s v_{\parallel}},\\
l^s_{55}&=&\left(e_s E_x\right)^2 l^s_{11}\left(\frac{1}{m_s v_{\parallel}}\right)^2.
\end{eqnarray}
\end{subequations}
And the related macroscopic transport matrix can be computed straightforwardly.
\begin{subequations}
\begin{eqnarray}
L^s_{\alpha 4}&=&e_s E_x L^s_{\alpha 1},\\
L^s_{\alpha 5}&=&e_s E_x n_s\left\langle l^s_{11}h^s_{\alpha}\frac{1}{m_s v_{\parallel}} \right\rangle,\\
L^s_{44}&=&\left(e_s E_x\right)^2 L^s_{11},\\
L^s_{45}&=&\left(e_s E_x\right)^2 n_s\left\langle  l^s_{11}\frac{1}{m_s v_{\parallel}} \right\rangle,\\
L^s_{55}&=&\left(e_s E_x\right)^2 n_s\left\langle  l^s_{11}\left(\frac{1}{m_s v_{\parallel}}\right)^2 \right\rangle.
\end{eqnarray}
\end{subequations}
Note that all the element of the transport matrix can be computed once $l^s_{11}=d^{xx}$ is known; this provides a useful scheme to evaluate the transport matrix in a kinetic turbulence numerical simulation where $d^{xx}$ can be computed in a straightforward way.

It is useful to explicitly write down the radial particle flux ($J^s_1$), the radial heat flux ($J^s_2$), and the radial component of the parallel viscosity ($J^s_3$).
\begin{subequations}
\begin{eqnarray}
J^s_{\alpha}&=&L^s_{\alpha \beta}A^s_{\beta}+J^s_{\alpha,res},\label{eq:Jn}\\
J^s_{\alpha,res}&=&\frac{e_s E_x}{T_s}\left(L^s_{\alpha 1}-n_s\left\langle l^s_{11}h^s_{\alpha}\frac{U_s}{v_{\parallel}}\right\rangle \right).\label{eq:Jr}
\end{eqnarray}
\end{subequations}
The first part of the residual particle flux ($J^s_{1,res}$), $\frac{e_s E_x}{T_s}L^s_{11}=n_s e_s E_x \frac{D_s}{T_s}$, due to the correlation between $\Delta x$ and $\Delta K$ induced by the REF, gives the usual contribution of particle species $s$ to the radial electrical conductivity, which agrees with the previous results for the case without any parallel flows \cite{MontgomeryPRL71}. This is the well-known Stokes-Einstein's formula of fluctuation dissipation \cite{Einstein1905a,PathriaBook12}. The second part, which is found for the first time, is due to the correlation between $\Delta x$ and $\Delta v_{\parallel}$ and non-zero for the case with finite parallel flow. These residual fluxes induced by the REF, in addition to the usual cross terms ($L^s_{\alpha \beta}A^s_{\beta}$, $\alpha \neq \beta$), also contribute to the pinch of particle, heat, and parallel momentum.

To understand the anomalous heat pinch, the turbulence heating rate given by Eq. (\ref{eq:t-h}) should be commented. The first term on the right-hand side of Eq. (\ref{eq:t-h}) represents the usual viscous heating, the second term ($\Gamma^x_s e_s E_x/T_s$) represents the Joule heating due to the radial electrical current, and the last term is $-F_s U_s/T_s$, which gives the consumption of the internal energy to accelerate the fluid.

If one considers the effect of the inhomogeneity of the equilibrium magnetic field in a tokamak, an additional change of kinetic energy of a deeply trapped particle induced by a random walk is $\Delta K=\mu B \partial_x \ln B \Delta x$. Using this $\Delta K$ that is again correlated with $\Delta x$, following the analysis in this paper one finds the pinch velocity of the deeply trapped particle $\frac{1}{T_s}\left\langle\frac{\Delta x \Delta K}{2\Delta t}\right\rangle_{\mathcal {E}}=\frac{\mu B}{T_s}\partial_x \ln B D_s$. Using $\mu B \sim T_s$ and $\partial_x \ln B\sim -1/R$ for the deeply trapped particle, with $R$ the major radius of the torus, one finds the radial pinch velocity $-D_s/R$ for the deeply trapped particle, which agrees with the previous results obtained by invoking the turbulence equipartition model \cite{IsichenkoPRL95,BakerPoP98}. Clearly, the main results of this paper do not change if one includes this effect.

The ambipolar REF $E_x$ is given by the ambipolarity condition, $e_i\Gamma^x_i+e_e\Gamma^x_e=0$.
For the case considered in Ref. \onlinecite{BakerPoP98} where the ion diffusivity is much larger than the electron diffusivity, the ambipolarity condition is reduced to $\Gamma^x_i\approx0$, which gives the ambipolar REF
\begin{equation}
E_x\approx-\frac{T_i}{e_i}\frac{L^i_{1 \alpha}A^i_{\alpha}}{L^i_{11}-n_i\left\langle l^i_{11}\frac{U_i}{v_{\parallel}}\right\rangle}. \label{eq:Ex}
\end{equation}
Clearly, only in the case with zero parallel velocity and zero temperature gradient the particle distribution satisfies the Boltzmann relation. For the case of weak flow, $U_i\ll v_{thi}$, with $v_{thi}$ the thermal ion speed, one can write

\begin{equation}
E_x\approx-\frac{T_i}{e_i}\frac{L^i_{1 \alpha}A^i_{\alpha}}{L^i_{11}}. \label{eq:Ex0}
\end{equation}

The residual parallel viscosity due to the ambipolar REF may contribute to the intrinsic rotation. To understand the parallel acceleration term due to the ambipolar REF, which is found for the first time, one can compare the following two equations.
\begin{equation}
\Gamma^x_{s}=n_s\left\langle l^s_{1i}A^s_{i}\right\rangle\equiv n_s\left\langle\gamma^x_s\right\rangle,
\end{equation}
\begin{equation}
F_s=e_s E_x n_s\left\langle\gamma^x_s \frac{1}{v_{\parallel}}\right\rangle. \label{eq:Fi}
\end{equation}
Clearly, the parallel acceleration is due to the fact that the work done by the REF in a random walk is used to increase the parallel kinetic energy, since the magnetic moment is conserved.

For the weak flow case, the ion momentum equation can be written as
\begin{equation}
\partial_t \left( n_i m_i U_i\right)+\partial_x \left(\Pi_{\parallel i}^x\right)=F_i,\label{eq:momentum-0}
\end{equation}
\begin{equation}
\Pi_{\parallel i}^x=\frac{T_i}{U_i}\left(L^i_{31}A^i_{1}+L^i_{32}A^i_{2}+L^i_{33}A^i_{3}+L^i_{31}\frac{e_i E_x}{T_i}\right),\label{eq:Pi-0}
\end{equation}
\begin{equation}
F_i\sim \Gamma^x_i e_i E_x m_i U_i/T_{i}, \label{eq:Fi-0}
\end{equation}
where in obtaining Eq. (\ref{eq:Fi-0}) from Eq. (\ref{eq:Fi}), we have used the Taylor expansion of the averaged distribution function around $v_{\parallel}=0$. Note that on the right-hand side of Eq. (\ref{eq:Pi-0}), the terms related to $A^i_{1}$, $E_x$ and $A^i_{2}$ are attributed to the residual Reynolds stress \cite{DiamondPoP08}.

For the typical tokamak experiment parameters \cite{RiceNF98}, the minor radius $a\sim 0.2m$, $n_i=n_e\sim 10^{20}/m^3$, $T_i\sim 2keV$, the deuteron thermal speed $v_{thi}\sim 3\times10^5m/s$, $U_i\sim 3\times10^4m/s$, $E_x \sim 30 kV/m $, the toroidal momentum confinement time $\tau_{\phi} \sim 0.1s$. If one assumes the particle confinement time is two times longer than $\tau_{\phi}$, one finds the particle flux $\Gamma_i^x =\Gamma_e^x \sim n_e \times 1.0m/s$. Using these parameters, one finds $F_i\sim 0.15 N/m^3$. The toroidal momentum dissipation rate is estimated as $n_i m_i U_i/\tau_{\phi} \sim 0.1 N/m^3$, which is comparable with the parallel acceleration rate due to the REF. This suggests that the parallel momentum acceleration due to the REF is a significant factor in understanding the intrinsic toroidal rotation in a tokamak.

(IV)SUMMARY

In conclusion, we have proposed a new theoretical model based on the Fokker-Planck phase-space transport equation to interpret the intrinsic rotation observed in tokamak experiments. The proposed model predicts a parallel acceleration of the plasma due to the REF through the correlation between the random walk in the radial direction ($\Delta x$) and the random change of the particle parallel velocity ($\Delta v_{\parallel}$). The REF, which should satisfy the ambipolarity condition, can be self-generated or externally applied. A radial step of random walk in the REF inevitably changes the kinetic energy of the charged particle; since the magnetic moment is conserved, a change of kinetic energy leads to a change of the parallel velocity of the particle; this explains the correlation between $\Delta x$ and $\Delta v_{\parallel}$. The connections of this theoretical model to the anomalous pinch \cite{IsichenkoPRL95,BakerPoP98} is also discussed. In contrast to the residual Reynolds stress model ($L^i_{31}\neq 0 \neq L^i_{32}$), which depends on the symmetry breaking effects \cite{DiamondPoP08}, the parallel acceleration term found here can not be written as a divergence. In contrast to the previous parallel acceleration mechanism \cite{WangPRL13}, which is the parallel acceleration rate induced by the correlation between the fluctuations of ion density and ion temperature, the new parallel acceleration mechanism is the parallel acceleration force induced by the correlation between $\Delta x$ and $\Delta v_{\parallel}$ due to the REF and magnetic moment conservation.

\begin{acknowledgments}
This work was supported by the National Natural Science Foundation of China under Grant No. 11175178, No. 11375196 and the National ITER program of China under Contract No. 2014GB113000.
\end{acknowledgments}

\nocite{*}


%

\end{document}